\documentclass[a4paper,12pt]{article}

\newcommand{\sect}[1]{\setcounter{equation}{0}\section{#1}}

\begin{document}
\topmargin 0pt \oddsidemargin 0mm

\renewcommand{\thefootnote}{\fnsymbol{footnote}}
\begin{titlepage}
\begin{flushright}
hep-th/0206223
\end{flushright}

\vspace{5mm}
\begin{center}
{\Large \bf Constant Curvature Black Hole and Dual Field Theory}
\vspace{12mm}

{\large
Rong-Gen Cai\footnote{Email address: cairg@itp.ac.cn}\\
\vspace{8mm}
{ \em Institute of Theoretical Physics, Chinese Academy of Sciences, \\
   P.O. Box 2735, Beijing 100080, China}}
\end{center}
\vspace{5mm} \centerline{{\bf{Abstract}}}
 \vspace{5mm}
We consider a five-dimensional constant curvature black hole,
which is constructed by identifying some points along a Killing
vector in a five-dimensional AdS space. The black hole has the
topology ${\cal M}_4\times S^1$, its exterior is time-dependent
and its boundary metric is of the form of a three-dimensional de
Sitter space times a circle, which means that the dual conformal
field theory resides on a dynamical spacetime. We calculate the
quasilocal stress-energy tensor of the gravitational background
and then the stress-energy tenor of the dual conformal field
theory. It is found that the trace of the tensor does indeed
vanish, as expected. Further we find that the constant curvature
black hole spacetime is just the ``bubble of nothing" resulting
from Schwarzschild-AdS black holes when the mass parameter of the
latter vanishes.

\end{titlepage}

\newpage
\renewcommand{\thefootnote}{\arabic{footnote}}
\setcounter{footnote}{0} \setcounter{page}{2}
\sect{Introduction}
 Recently there has been much interest in studying string theory
 in time-dependent spacetimes. The authors of papers~\cite{strings}
  discussed orbifold constructions giving solutions with tractable
  string descriptions.  Aharony {\it et al}~\cite{Aharony} considered
  spacetimes of so-called ``bubbles of nothing", which are double
  analytic continuations of Schwarzschild or Kerr black hole
  spacetimes. These spacetimes are interesting examples of smooth
  time-dependent solutions because they are consistent backgrounds
  for string theory at least to leading order since they are
  vacuum solutions to Einstein's equations.  More recently, Balasubramanian
  and Ross~\cite{BR} have extended the discussions of
  \cite{Aharony} to the cases of asymptotically anti-de Sitter
  (AdS) spacetimes. They considered ``bubbles of nothing"
  consturcted by analytically continuing (Schwarzschild, Reissner-Nordstr\"om,
  and Kerr) black holes in AdS spaces. Since these spacetimes are
  asymptotically AdS, so it is possible to study them through the
  AdS/CFT correspondence~\cite{AdS}. Further since these
  spacetimes are time-dependent and then the boundary metric is also
  time-dependent, in some sense it therefore opens a window to
  study dual strong coupling field theory in time-dependent
  backgrounds. For the ``bubbles of nothing" resulted from double
  analytic continuation of the Schwarzschild-AdS black hole
  spacetime, its boundary metric is of the form of de Sitter (dS)
  space times a circle. Balasubramanian and Ross calculated
  stress-energy tensor of dual field theory to ``bubbles of
  nothing". The bubble solutions in AdS spaces are first
  constructed in \cite{Birm}.

 The work of Balasubramanian and Ross is reminiscent of the
 constant curvature black holes constructed by
 Banados~\cite{Bana1} by identifying some points along a Killing
 vector in an AdS space. The so-called constant curvature black
 holes are analogues of BTZ black holes in higher dimensions. The
 topology structure of the black hole are quite different from that
  of usual black holes. For example, in $D$ dimensions the
 usual black holes have the topology ${\cal M}_2 \times S^{D-2}$,
 while the constant curvature black holes have the topology
 ${\cal M}_{D-1}\times S^1$. Here $M_n$ denotes a conformal
 Minkowski space in $n$ dimensions. Of course, in higher dimensions
 ($D>4$) it is possible to have  black hole structure
 if one replaces the factor $S^{D-2}$ by other topologies in ${\cal
 M}_2\times S^{D-2}$. In the asymptotically AdS space, even in four
 dimensions one can have other topological even horizons by replacing
 $S^2$ by other forms, resulting in so-called topological black
 holes (for a more or less complete list of references see~\cite{GB}).
 The causal structure of those black holes is
  still determined by
 the two-dimensional manifold ${\cal M}_2$ for these cases.
 For the constant curvature black holes, however, the causal
 structure is determined by a $(D-1)$-dimensional manifold ${\cal
 M}_{D-1}$. Of special interest is that the exterior of these black
 holes are time-dependent. So it is possible to discuss dual field
 theory in these dynamical spacetimes through the AdS/CFT
 correspondence.

 In this note we consider a five-dimensional constant curvature
 black hole. This is of particular interest because its dual is
 ${\cal N}=4$ supersymmetric Yang-Mills theory if the AdS/CFT
 correspondence holds in the case we are discussing. In Sec.~2
 we introduce the construction of the black hole and
 discuss some salient features of the black hole~\cite{Bana1,Bana2}.
  Although the black hole has the strange topology
 and its exterior is not static, we find that the surface counterterm
 approach~\cite{BK} still works well. In Sec.~3 using the
 surface counterterm approach we calculate the quasilocal
 stress-energy tensor of gravitational field and identify it to
 the stress-energy tensor of  the dual field theory.  Sec.~4 is devoted to
 comparing with related investigations
 to the black hole by other authors, and to discussing the relation
 between our results and those in \cite{BR}. In particular we
 point out that the constant curvature black hole spacetime is
 just the ``bubble of nothing" resulting from the
 Schwarzschild-AdS black hole if $r_0=0$ in~\cite{BR}.

\sect{Constant curvature black holes}

A five dimensional AdS space is defined as the universal covering
 space of a surface obeying
\begin{equation}
\label{2eq1}
 -x_0^2 +x_1^2 +x_2^2 +x_3^2 +x_4^2-x_5^2=-l^2,
\end{equation}
where $l$ is the AdS radius. This surface has fifteen  Killing
vectors, seven rotations and eight boosts. Consider the boost
$\xi=(r_+/l)(x_4\partial_5+x_5\partial_4)$ with norm
$\xi^2=r_+^2(-x_4^2+x_5^2)/l^2$, where $r_+$ is an arbitrary real
constant. The norm can be negative, zero, or positive. In terms of
the norm, the surface (\ref{2eq1}) can be re-expressed as
\begin{equation}
\label{2eq2}
 x_0^2=x_1^2 +x_2^2 +x_3^2 +l^2(1-\xi^2/r_+^2).
 \end{equation}
 When $\xi^2=r_+^2$, the surface (\ref{2eq2}) reduces to a null
 one
 \begin{equation}
 \label{2eq3}
 x_0^2=x_1^2+x_2^2 +x_3^2,
 \end{equation}
 while $\xi^2=0$, it becomes a hyperboloid
 \begin{equation}
 \label{2eq4}
x_0^2=x_1^2+x_2^2 +x_3^2 +l^2.
\end{equation}
The cone (\ref{2eq3}) has two pointwise connected branches, called
$H_f$ and $H_p$~\footnote{Ref.~\cite{Bana2} discussed the four
dimensional case, it is straightforward to generalize to the
higher dimensional cases.}, defined by
\begin{eqnarray}
 H_f:&& x^0=+\sqrt{x_1^2 +x_2^2 +x_3^2}, \nonumber\\
 H_p:&& x^0=-\sqrt{x_1^2 +x_2^2 +x_3^2}.
 \end{eqnarray}
Similarly, the hyperboloid (\ref{2eq4}) has two disconnected
branches, named $S_f$ and $S_p$,
\begin{eqnarray}
 S_f:&& x^0=+\sqrt{x_1^2 +x_2^2 +x_3^2+l^2}, \nonumber\\
 S_p:&& x^0=-\sqrt{x_1^2 +x_2^2 +x_3^2+l^2}.
 \end{eqnarray}
The Killing vector $\xi$ is spacelike in the region contained
in-between $S_f$ and $S_p$, is null at $S_f$ and $S_p$ and is
timelike in the causal future of $S_f$ and in the causal past of
$S_p$.

Identifying the points along the orbit of $\xi$, another
one-dimensional manifold becomes compact and isomorphic to $S^1$.
The region where $\xi^2 <0$ has a pathological chronological
structure and therefore it must be cut off from physics spacetime.
In this sense, the surface $\xi^2=0$ is a singularity, $S_f$ is
the future one and $S_p$ the past one. The surface $\xi^2=r_+^2$
is a horizon, $H_f$ is the future one and $H_p$ the past.  Through
the above analysis, it turns out that one can construct a black
hole by identifying point along the orbit of the Killing vector
$\xi$. Since the starting point is the AdS, the resulting black
hole therefore has a constant curvature as the AdS. The topology
of the black holes is ${\cal M}_{4}\times S^1$, which is quite
different from the usual topology, ${\cal M}_2\times S^3$, of
five-dimensional black holes, where ${\cal M}_n$ denotes a
conformal Minkowski space in $n$ dimensions. The Penrose diagram
of the constant curvature black holes has been drawn in
Refs.\cite{Bana1,Bana2}.

The constant curvature black holes can be best described by using
Kruskal coordinates. In Ref.\cite{Bana1} a set of coordinates on
the AdS for the region $\xi^2 >0$ has been introduced. The six
dimensionless local coordinates $(y_i,\varphi)$ are
\begin{eqnarray}
\label{2eq7}
 && x_i=\frac{2ly_i}{1-y^2}, \ \ \ i=0,1,2,3 \nonumber \\
 && x_4=\frac{lr}{r_+}\sinh\left(\frac{r_+\varphi}{l}\right),
 \nonumber \\
 && x_5=\frac{lr}{r_+}\cosh\left(\frac{r_+\varphi}{l}\right),
 \end{eqnarray}
 with
 \begin{equation}
 r=r_+\frac{1+y^2}{1-y^2}, \ \ \ y^2=-y_0^2+y_1^2+y_2^2+y_3^2.
 \end{equation}
Here $ -\infty <y_i < \infty$ and $-\infty <\varphi <\infty$ with
the restriction $-1 <y^2 <1$. In these coordinates the boundary
$r\to \infty$ corresponds to the hyperbolic ``ball" $y^2=1$, and
the induced metric can be written down
\begin{equation}
\label{2eq8}
 ds^2 = \frac{l^2(r+r_+)^2}{r_+^2}(-dy_0^2
+dy_1^2+dy_2^2+dy_3^2)
    +r^2 d\varphi^2.
\end{equation}
Obviously the Killing vector is $\xi=\partial_{\varphi}$ with norm
$\xi^2=r^2$. The black hole spacetime is thus simply obtained by
identifying $\varphi \sim \varphi +2\pi n$ and the topology of the
black hole clearly is ${\cal M}_4\times S^1$.

The constant curvature black holes can also be described by using
Schwarzschild coordinates. Introducing local ``spherical"
coordinates ($t,r,\theta,\chi$) in the hyperplane $y_i$
\cite{Bana1}:
\begin{eqnarray}
\label{2eq10}
 && y_0=f \cos\theta \sinh(r_+t/l),\ \ \
y_1=f\cos\theta
    \cosh(r_+t/l), \nonumber \\
 && y_2=f \sin\theta \sin\chi, \ \ \ \ \ \ y_3=f \sin\theta \cos\chi,
\end{eqnarray}
where $f=[(r-r_+)/(r+r_+)]^{1/2}$, $0\le \theta \le \pi/2$, $0\le
\chi \le 2\pi$ and $r_+ \le r< \infty$, one can find that the
solution (\ref{2eq8}) becomes
\begin{equation}
\label{2eq11}
 ds^2= l^2 N^2 d\Omega_3 +N^{-2}dr^2 +r^2d\varphi^2,
 \end{equation}
 where
 \begin{equation}
 N^2=\frac{r^2-r_+^2}{l^2}, \ \ \ d\Omega_3=-\sin^2\theta dt^2
 +\frac{l^2}{r_+^2}(d\theta^2 +\cos^2\theta d\chi^2).
 \end{equation}
 This is the black hole solution in the Schwarzschild coordinates.
 Here $r=r_+$ is the black hole horizon location.
 In these coordinates the solution looks static. But we can see
 from (\ref{2eq10}) that the form (\ref{2eq11}) does not cover
 the full outer region of black hole since the difference
 $y^2_1-y_0^2$ is constrained to be positive in the region covered
 by these coordinates. Indeed, it has been proved that there is no
 globally timelike Killing vector in this geometry~\cite{Holst}.

Similar to the case in four dimensions~\cite{Bana2}, we find that
there is another set of coordinates, which has the advantage of
covering the entire exterior of the Minkowskian black hole
geometry:
\begin{eqnarray}
\label{2eq13}
 && y_0 = f\sinh(r_+t/l), \ \ \ y_1= f\cos\theta \cosh(r_+t/l),
 \nonumber \\
  && y_2= f\sin\theta \cos\chi \cosh(r_+t/l), \ \ \
  y_3=f\sin\theta\sin\chi \cosh(r_+t/l),
  \end{eqnarray}
  where $f$ is given as before, $ 0\le \theta \le \pi$, $r_+ \le
  r<\infty$ and $0 \le \chi \le 2\pi $. In terms of these
  coordinates, the solution can be expressed as
  \begin{equation}
  \label{2eq14}
  ds^2=N^2l^2 d\Omega_3+ N^{-2}dr^2 +r^2d\varphi^2,
  \end{equation}
  where $N^2=(r^2-r_+^2)/l^2$ and
  \begin{equation}
  \label{2eq15}
  d\Omega_3=-dt^2 +\frac{l^2}{r_+^2} \cosh^2(r_+t/l)(d\theta^2
  +\sin^2\theta d\chi^2).
  \end{equation}
 In these coordinates, the time-dependence of the solution
  is obvious.
 This situation is quite similar to the case of de Sitter
 spacetimes, where in the static coordinates the de Sitter space
 looks static within the cosmological horizon, but not cover the
 whole
 de Sitter space, while in the global coordinates the solution
 covers the whole space, but is obviously time-dependent. A
 different point is that here both sets of coordinates
 (\ref{2eq10}) and (\ref{2eq13}) only describe the exterior of
 the black hole.

 The Euclidean black hole solution can be obtained by replacing
 $t$ by $ -i( \tau + \pi l /(2 r_+))$ in (\ref{2eq15}). In this case,
 $d\Omega_3$ becomes
 \begin{equation}
 \label{2eq16}
 d\Omega_3=d\tau^2
 +\frac{l^2}{r_+^2}\sin^2(r_+\tau/l)(d\theta^2+\sin^2\theta d\chi^2).
 \end{equation}
In order the $\Omega_3$ to be a regular three-sphere, the $\tau$
must have the range, $0 \le \tau \le \beta $ with
\begin{equation}
\label{2eq17}
 \beta =\frac{\pi l}{r_+}.
 \end{equation}
 Regarding this as the inverse Hawking temperature of the black hole seems
 problematic since the surface gravity of the black hole is $\kappa =r_+/l$
 and $\beta$ does not obey the usual relation of black hole thermodynamics,
 $\beta = 2\pi/\kappa$. Needless to say, it is of great interest
 to further study the Hawking evaporation of the black hole.

 Defining $r^2=r_+^2(1+\tilde r^2/l^2)$, and $\tau =l\phi /r_+$,
 we find that the Euclidean black hole becomes
 \begin{equation}
 \label{2eq18}
 ds^2=\left(1+\frac{\tilde r^2}{l^2}\right)d(r_+\varphi)^2
    +\left(1+\frac{\tilde r^2}{l^2}\right)^{-1}d\tilde r^2
    +\tilde r^2(d\phi^2 +\sin^2\phi (d\theta^2 +\sin^2\theta
    d\chi^2 )).
 \end{equation}
 Since the $\tau$ has the range from zero to $\beta$, so we have $0\le \phi
 \le \pi$. The solution (\ref{2eq18}) is a Euclidean AdS space if
 one regards $ r_+ \varphi $ as the Euclidean time. Because the
 $\varphi$ has the period $2\pi$, the solution (\ref{2eq18}) can
 be viewed as a thermal AdS space~\cite{Bana2}. This is the
 Euclidean version of the black hole.

The vacuum state is subtle for the black hole solution
(\ref{2eq14}) because it has not a smooth limit as $r_+ \to 0$.
Redefining $\theta =r_+\tilde \theta /l$, one can obtain a
well-defined limit of the solution (\ref{2eq14}):
\begin{equation}
\label{2eq19}
 ds^2 =\frac{l^2}{\rho^2}\left( -dt^2 + d\rho^2 +d\tilde \theta^2
 +\tilde\theta^2 d\chi^2 +l^2 d\varphi^2\right),
 \end{equation}
 where $\rho =l/r$ and the range of $\tilde \theta$ is $0\le
 \tilde \theta < \infty$. The form of the solution (\ref{2eq19})
 looks like the one of AdS space in the Poincare coordinates,
 however, the coordinate $\varphi$ has a period $2\pi$. It implies
 that the form (\ref{2eq19}) is an AdS space with identified
 points. It is of course right because the constant curvature
 black hole is obtained by identifying some points in an AdS
 space.

\sect{Dual field theory: Stress-energy tensor}

Since the constant curvature black hole (\ref{2eq14}) is obtained
by identifying some points along a Killing vector in an AdS space,
so one can regard the black hole spacetime as a solution of
Einstein's equations with a negative cosmological constant, namely
 a solution of the following action:
\begin{equation}
\label{3eq1}
 S=\frac{1}{16\pi G} \int_{\cal M} d^5x\sqrt{-g}\left({\cal R}+
   \frac{12}{l^2}\right) -\frac{1}{8\pi G}\int_{\partial \cal M}
   d^4x\sqrt{-h}K,
  \end{equation}
 where the second term is the Hawking-Gibbons surface term, $K$ is
 the trace of the extrinsic curvature for the
 boundary $\partial {\cal M}$ and $h$ is the induced metric of the
 boundary. For a five-dimensional asymptotic AdS space, the
 suitable surface counterterm is~\cite{BK}
 \begin{equation}
 S_{\rm ct}=-\frac{1}{8\pi G} \int_{\partial \cal
       M}d^4x\sqrt{-h}\left(\frac{3}{l}+\frac{l}{4}R\right),
  \end{equation}
  where $R$ is the curvature scalar for the induced spacetime $h$
  and $\cal R$ for the bulk spacetime $g$. With this counterterm,
  the quasilocal stress-energy tensor of the gravitational field
  is
  \begin{equation}
  T_{ab}=\frac{1}{8\pi G}\left (K_{ab}-Kh_{ab}-\frac{3}{l}h_{ab}
      +\frac{l}{2}G_{ab}\right),
   \end{equation}
   where $G_{ab}$ is the Einstein tensor for the induced metric,
   $K_{ab}=-h_{a}^{\ \mu} \nabla_{\mu}n_b$, and $n_a$ is the unit
   normal to the boundary surface.

  For the black hole spacetime (\ref{2eq14}), we choose a timelike
  hypersuface with a fixed $r(>r_+)$. Calculating the extrinsic
  curvature and the Einstein tensor for the induced metric $h$,
  we obtain
  \begin{eqnarray}
  \label{3eq4}
  && T_{tt}=-\frac{1}{8\pi G l}\frac{r_+^4}{8r^2}+\cdots,
     \nonumber \\
  &&T_{\theta\theta}=\frac{l \cosh^2(r_+t/l)}{8\pi
     G}\frac{r_+^2}{8r^2}+\cdots, \nonumber \\
  &&T_{\chi\chi}=\frac{l \cosh^2(r_+t/l)\sin^2\theta}{8\pi
     G}\frac{r_+^2}{8r^2}+\cdots, \nonumber \\
   && T_{\varphi\varphi} =-\frac{1}{8\pi G l}\frac{3r_+^4}{8r^2}
      +\cdots,
  \end{eqnarray}
where $\cdots $ denote higher correction terms, which will vanish
when we take the limit $r\to \infty$. For the black hole
spacetime, up to a conformal factor the boundary metric, on which
the dual field theory resides, is,
\begin{equation}
\label{3eq5}
 ds^2_{\rm CFT}=\gamma_{ab}dx^adx^b=
   -r_+^2 dt^2 +l^2\cosh^2(r_+t/l)(d\theta^2
  +\sin^2\theta d\chi^2) +r_+^2 d\varphi^2.
  \end{equation}
 Note that here the coordinate $t$ is dimensionless.
  This spacetime is obviously a $dS_3\times S^1$, which is a
  time-dependent background.
Corresponding to the boundary metric (\ref{3eq5}), the
stress-energy tensor for the dual field theory can be calculated
using the following relation~\cite{Myers}
\begin{equation}
 \label{3eq6}
  \sqrt{-\gamma}\gamma^{ab}\tau_{bc}=\lim_{r\to
  \infty}\sqrt{-h}h^{ab}T_{bc}.
  \end{equation}
  Substituting (\ref{3eq4}) into (\ref{3eq6}), we obtain
  \begin{eqnarray}
  \label{3eq7}
  && \tau_{t}^{\ t}=\frac{1}{8\pi G}\frac{1}{8l}, \ ~~~~~~~~~
 \tau_{\theta}^{\ \theta}=\frac{1}{8\pi G}\frac{1}{8l},
     \nonumber \\
   && \tau_{\chi}^{\ \chi}=\frac{1}{8\pi G}\frac{1}{8l}, \ ~~~~~~~~
      \tau_{\varphi}^{\ \varphi}=-\frac{1}{8\pi
   G}\frac{3}{8l}.
   \end{eqnarray}
   Some points about this stress-energy tensor are worthwhile to
   mention here. First of all, we notice that the tensor is independent
   of the parameter $r_+$. This can be understood since the black hole
   solution is locally equivalent to the AdS space. That is, after a
   rescaling of coordinates the dependence of the metric on $r_+$ can
   disappear.  Second, although the tensor is the function of the
   radius $l$ of AdS space only, it is not the contribution of the
   Casimir effect (in the case of five-dimensional Schwarzschild-AdS
   black holes the stress-energy tensor of the dual CFT has a Casimir
   term~\cite{BK}).  There is
   no Casimir energy associated with the black hole solution
   (\ref{2eq14}). This can be understood from the vacuum
   state (\ref{2eq19}), an AdS space in the Poincare coordinates
   with some identification. It is well-known that there is no
   Casimir energy for the AdS black holes with Ricci flat
   horizons~\cite{Cai2}, which also holds even for asymptotically
   de Sitter spaces~\cite{Cai3}.
   Third, the trace of the tensor does vanish, as expected, since
   dual field theory is the ${\cal N}=4$ supersymmetric Yang-Mills
   theory, whose conformal symmetry is not broken even if quantum
   corrections are taken into account. Further, for the conformal
   field theory in four dimensions, in general there is a
   conformal anomaly proportional
    to $(R_{ab}R^{ab}-R^2/3)$~\cite{BR}, where
   $R_{ab}$ and $R$ are the Ricci tensor and curvature scalar of the
   boundary metric,  respectively. In our case,  it is easy to check that
    this term vanishes for
   the background (\ref{3eq5}). Finally, the positive sign  of
   the quantity $\tau_t^{\ t}$ indicates that the black hole has a negative mass
   and the dual field theory has a negative energy density.

 \sect{Discussions}

  A $D$-dimensional constant curvature black hole has unusual topological
  structure ${\cal M}_{D-1}\times S^1$. So it is quite difficult
  to calculate some conserved quantities associated with the black
  hole spacetime, in particular due to that there is no globally
  timelike Killing vector in the geometry of the black hole, which is quite similar
  to the case of asymptotically de Sitter spaces. In \cite{Bana1}
  Banados considered a five dimensional rotating constant
  curvature black hole and embedded it to a Chern-Simons
  supergravity theory. By computing related conserved charges it
  was found that the black hole mass is proportional to the
  product of outer horizon $r_+$ and inner horizon $r_-$, while
  the angle momentum proportional to the sum of two horizons.
  Further the entropy of black hole is found to be proportional to
  the inner horizon $r_-$.  This
  approach has two obvious drawbacks. Firstly the result cannot be
  degenerated to the non-rotating case. Secondly it cannot be
  generalized to other dimensional cases.

   In \cite{Mann} Creighton and Mann considered the quasilocal
   thermodynamics of a four dimensional constant curvature black
   hole in general relativity by calculating some thermodynamic
   quantities at a finite boundary which encloses the black hole.
   In this context it was found that the entropy is not associated
   with the event horizon, but the Killing horizon of a static
   observer, which is tangent to the event horizon of the black
   hole. The quasilocal energy density (see (11) in \cite{Mann})
   is  found to be negative.

   In this paper we considered a five-dimensioanl constant
   curvature black hole. The exterior of the black hole is time-dependent
   and the boundary spacetime is of the form of a $dS_3$ times a circle.
   By employing the surface counterterm approach, we obtained quasilocal
   stress-energy tensor of gravitational field and then that of the dual
   conformal field theory which resides on the dynamical boundary spacetime.
   As expected, the trace of the tensor vanishes since the dual field theory
   is ${\cal N}=4$ super Yang-Mills theory. It is found that the
   field theory has a negative energy density on the
   time-dependent background. In addition, we would like to emphasize that
    the surface counterterm approach does not apply in four or six dimensions,
   but we do not know whether it applies in seven dimensions or more
   higher dimensions at the moment~\footnote{ In the earlier version of this
   paper, we also discussed the thermodynamics of the black hole through
   calculating the Euclidean action of black hole with the surface counterterm
   approach. Both the mass and entropy were found to be negative.
   However, as pointed out to us by the referee, the interpretation of the $\beta$ as
   the inverse Hawking temperature of the black hole seems problematic.
   As a result, the resulting calculations concerning the mass and entropy become
   unbelievable. We thank the referee for pointing this out.}.

   Since our boundary metric is a $dS_3\times S^1$ spacetime,
   the same as
   the one for the ``bubbles of nothing" in \cite{BR}, a natural
   question then arises: what is the relation between our discussion
   and \cite{BR}? We notice that the spacetime we considered
   is a black hole solution, the dual field theory is then at finite
   temperature and the parameter $r_+$ is related to the
   Hawking temperature of the black hole, while the ``bubble
   of nothing" is not a black hole solution and the physical meaning of
   the parameter $r_0$ in \cite{BR} is not very clear. However, we find
   that when $r_0=0$ the solution (2) in \cite{BR} can be
   identified with the constant curvature black hole (\ref{2eq14})
   if $r_+=l$ through a suitable coordinate transformation.
   Indeed, in this case it can be seen that the stress-energy
   tensor (18) of dual field theory in \cite{BR} is completely the
   same as ours [see (\ref{3eq7})]. Here it should be pointed out
   that when $r_0 \neq 0$ the ``bubble of nothing" solution in
   \cite{BR}  cannot be explained as a black hole. Further we
   would like to mention that the vacuum solutions are different
   in both cases, in our case the vacuum solution is
   (\ref{2eq19}), an AdS space in Poincare coordinates with some
   identifying points, while the vacuum solution  is
   the one (2) with $r_0=0$ in \cite{BR}. This might give rise to
   some differences in explaining the stress-tensor tensor of dual
   field theory.

   It would be interesting to further study some properties of
   dual field theory to the constant curvature black hole through
   the AdS/CFT correspondence, for example to calculate the Wilson
   loop.

\section*{Acknowledgments}
The author thanks J.X. Lu and Y.S. Wu for quite helpful
discussions, Z.Y. Zhu for a reading of manuscript, Balasubramanian
and Ross for useful comments and for pointing out a calculation
error in the earlier version. This work was supported in part by a
grant from Chinese Academy of Sciences.

\end{document}